\documentclass[aps,pra,reprint,
nofootinbib,
amsmath,amssymb,
aps,amsthm,
]{revtex4-2}
\usepackage{hyperref}

\begin{document}
	\title{Wave Function Collapse, Lorentz Invariance, and the Third Postulate of Relativity}
	
	\author{Edward J. Gillis}
	\email[]{gillise@provide.net}
	
	\noaffiliation
	
	\date{\today}

	\begin{abstract}

		\noindent 
	
		The changes that quantum states undergo during measurement are both probabilistic and nonlocal. These two characteristics complement one another to insure compatibility with relativity and maintain conservation laws. The probabilistic nature of nonlocal effects prevents the superluminal transmission of information,  while nonlocal entanglement relations provide a means to enforce conservation laws in a probabilistic theory.  In order to explain measurement-induced changes in terms of fundamental physical processes it is important to take these two key characteristics into account. One way to do this is to modify the Schr\"{o}dinger equation by adding stochastic, nonlinear terms. A number of such proposals have been made over the past few decades. A recently proposed equation based on the assumption that wave function collapse is induced by a sequence of correlating interactions of the kind that constitute measurements has been shown to maintain strict adherence to conservation laws in individual instances, and has also eliminated the need to introduce any new, ad hoc physical constants.

		In this work it is shone that the proposed stochastic modification to the Schr\"{o}dinger equation is also Lorentz invariant, even though it is formulated in a preferred reference frame. It is further argued that the additional spacetime structure that the proposed modification  requires provides a way to implement the assumption that spacelike-separated operators (and measurements) commute, and that this assumption of local commutativity should be regarded as a third postulate of relativity.

	\end{abstract}

	
	\maketitle

 \section{Introduction}

 If quantum theory is regarded as an objective description of the physical world then it should be possible, at least in principle, to explain how individual measurement outcomes are generated from fundamental processes. Because quantum states change in a probabilistic and nonlocal manner during measurements it is reasonable to suppose that these features will play key roles in constructing such a fundamental explanation. One major approach to this issue takes these features into account by adding stochastic, nonlinear terms to the Schr\"{o}dinger equation. These additional terms are designed to induce the wave function to collapse to one of its several branches. The nonlinearity is necessary in order to generate collapse, and the stochasticity is required in order to prevent superluminal signaling, as shown in a work by Gisin\cite{Gisin_c}.

 Gisin's work was one of a number of proposed stochastic modifications of the Schr\"{o}dinger equation aimed at resolving the measurement problem\cite{Pearle_1976,Pearle_1979,Gisin_1984,GRW,Diosi_1,Diosi_2,Diosi_3,GPR,GP_1,GP_2,Adler_Brun, Ghirardi_Bassi,Pearle_1,Brody_finite}. Most of these proposals are designed to collapse the state vector to either an approximate position state or to an energy eigenstate. These attempted solutions have often been met with skepticism because they introduce new, ad hoc physical constants and imply small violations of conservation laws. A recent work has shown how to eliminate these problematic  features\cite{Gillis_JPA1}. 
 It is based on the idea that wave function collapse is induced by the elementary interactions that establish correlations between physical systems, and was motivated by the fact that these correlating interactions play a central role both in the measurement of physical quantities and in the instantiation and transmission of physical information. In this work it will be shown that the proposed modification to the equation is Lorentz invariant.\footnote{Of course, the nonrelativistic Schr\"{o}diger equation is not, itself, invariant. What will be shown here is that the stochastic modification to the equation \textit{is} Lorentz invariant.}

The fact that wave function collapse can be described in a Lorentz invariant manner calls into question our current understanding of relativity. What do these nonlocal, but Lorentz invariant effects imply about the ontology of spacetime?

 Concerns about nonlocality were first raised almost immediately after the Schr\"{o}dinger equation was proposed (by Einstein at the 1927 Solvay conference; see also \cite{EPR}.) However, they were really sharpened by Bell when he showed that the correlations between entangled systems that are separated by spacelike intervals could not be explained by any account in which all physical processes are restricted to propagate only within the light cone\cite{Bell_1}. At least in some sense, measurements do affect spacelike-separated systems.

 The nonlocal nature of these effects has been, by far, the biggest challenge to developing a fully satisfactory account of quantum measurement because of the \textit{apparent} conflict with relativity. This challenge has often seemed insurmountable because of the extreme reluctance to consider the possibility that the nonlocal correlations implied by quantum theory might require that we modify or supplement the metric structure of relativistic spacetime (which is based on classical physics).

 The reason that the nonlocal effects do not generate any manifest conflicts with relativity is that they are fundamentally probabilistic. More specifically they obey the Born probability rule\cite{Born}. The rule was discovered empirically and was simply tacked on to quantum theory in an ad hoc manner. There was no serious effort to integrate it into the mathematical structure of the theory; nor was it immediately associated with relativity. 
 
 Relativistic quantum field theory deals with the nonlocal correlations in a somewhat more formal manner. It assumes that spacelike-separated operators commute (or anticommute), and this assumption implies the Born rule, thus preserving the Lorentz invariance of the theory. In his text on quantum field theory Weinberg is quite explicit about the fact that this is an \textit{additional} assumption that is necessary to maintain the relativistic character of the theory when he states that the assumption of the commutativity of spacelike-separated operators is made in order to preserve the Lorentz invariance of the scattering matrix\cite{Weinberg}. He specifically states that he is not linking local commutativity with the notion of causality. It appears that his motivation for emphasizing this point is that `causality' is used \textit{both} as a synonym for the no-superluminal-signaling principle, and a shorthand for the idea that no physical processes can propagate faster than light. He seems to rightly regard the conflation of these two concepts as a mistake.

The critical point is that the assumption of local commutativity functions effectively as a third postulate for relativity. Nevertheless, very little consideration has been given to the possible implications this postulate has for spacetime structure. The reasons for the extreme reluctance are fairly obvious. The conventional picture of relativistic spacetime is very elegant, and it beautifully captures our intuitive notion that causal processes propagate through space in a continuous manner. Historically, both special and general relativity preceded the full development of quantum theory, and ideas about spacetime had become pretty firmly fixed by the time the Heisenberg and Schr\"{o}dinger  equations were published\cite{Heis_Eqn,Schr_Eqn}.

The problem is that this reluctance has left the logical structure of contemporary physics in a very muddled state. The rules governing individual measurement outcomes imply a type of change at odds with the unitary evolution described by the  Schr\"{o}dinger equation, and there is no clear definition of the range of applicability of the two distinct types of change.

The stochastic nonlinear modifications of the Schr\"{o}dinger equation mentioned above do explain how individual measurement outcomes are selected in accord with the Born rule, and thereby provide a more complete and unified mathematical framework for contemporary physics. Since these modifications require adding structure to relativistic spacetime, let us consider the reasons for such a move.

In the conventional view relativistic spacetime is a four-dimensional manifold with a Lorentzian metric that defines a light cone structure. This framework implements Einstein's two postulates for relativity by prohibiting the assignment of an absolute temporal order to events that are separated by a spacelike interval. Why, then, did the advent of quantum theory necessitate the introduction of a third postulate to maintain this prohibition? Prior to the development of quantum theory the relativistic prohibition on temporal order was typically associated with the presumption that no physical processes could propagate outside the light cone. But the nonlocal correlations implied by quantum theory strongly suggest that there are physical effects that propagate across spacelike intervals. To ``explain" why these effects do not result in any manifest inconsistencies with relativity the new theory simply ruled by fiat that they had to respect the prohibition on temporal ordering. As noted, this was done by requiring that spacelike-separated operators commute. For the reasons mentioned a few paragraphs back there was no inclination to consider the possibility that spacetime possessed additional structure. In fact, the assumption of local commutativity was often conflated with the limitation on the speed of light. The following quote from Gell-Mann, Goldberger, and Thirring illustrates this point\cite{Gell-Mann_G_Th}:
\begin{quote}
	``The quantum mechanical formulation of the demand that waves do not propagate faster than the speed of light is, \textit{as is well known}, the condition that the measurement of two observable physical quantities should not interfere if the points of measurement are space-like to each other...the commutators of two Heisenberg operators... shall vanish if the operators are taken at space-like points." 
	(italics added)
\end{quote}  
(This quotation was cited by Bell in his discussion of local commutativity\cite{Bell_LNC}).

As suggested by Weinberg's very careful characterization of local commutativity described above, the kind of conflation demonstrated in the quotation is simply wrong. The assumption that spacelike-separated operators commute is made to insure that any effects on physical systems that \textit{do} propagate across spacelike intervals do not transmit any \textit{physical} information. It involves the (at least) implicit recognition that there are nonlocal effects, and that there is a need to regulate them. If this were not the case there would be no need to make an additional assumption. But, this kind of regulation ought to be explained, and not simply imposed by fiat.

As mentioned above Einstein's postulate about the invariance of the speed of light is implemented by attributing a light cone structure to spacetime. In other words, this postulate is \textit{explained} as a consequence of the fundamental nature of spacetime.  Should we not then also consider adding structure to spacetime to explain how nonlocal effects propagate and how they can be regulated to maintain Lorentz invariance? This is exactly what stochastic collapse equations do. By assuming a foliation of spacetime into spacelike surfaces, and invoking a stochastic process (or processes) they provide the desired explanation. By making wave function collapse and the Born rule follow from the fundamental equation of the theory, they provide a more coherent logical structure
for contemporary physics.

In order to provide a fully coherent logical framework for contemporary theory it is crucial that any proposed modifications to the fundamental equations be Lorentz invariant. The goals of this work are to show that the proposal of \cite{Gillis_JPA1} meets this criterion, and to examine its implications for our understanding of relativity. The next section describes the basic assumptions employed by nonlinear, norm-preserving stochastic collapse equations and illustrates how they work. Section III reviews the modification proposed in \cite{Gillis_JPA1}, showing how it eliminates the need to introduce new ad hoc physical constants and insures that conservation laws are respected \textit{in individual experiments}. Section IV demonstrates that the stochastic modification described in the previous section is Lorentz invariant, despite being formulated in a nonrelativistic framework. Based on the demonstration of Lorentz invariance Section V examines the implications for our understanding of  relativity and spacetime ontology.

\section{Stochastic Collapse Equations}

This section first provides an explanation of why stochastic collapse equations need to add structure to spacetime, specifically a preferred foliation associated with a stochastic process (or processes). This is followed by an illustration of how these dynamic equations generate collapse in accord with the Born rule. The literature cited earlier contains more general and formal demonstrations of how these equations work; a proof that the equation described in Section III entails collapse in conformity to the Born rule has been presented in \cite{Gillis_JPA1}. The purpose here is just to provide an intuitive understanding based on a simplified case that captures the essential features. The idea is to show that it is \textit{not} necessary to paste the measurement postulates onto quantum theory in an ad hoc manner, but rather that they follow in a very natural way from a relatively simple modification of the Schr\"{o}dinger equation that takes into account the fundamentally probabilistic nature of quantum theory as we currently understand it.

\subsection{The Need for Additional Spacetime Structure}

Interactions play a critical role in stochastic collapse equations. In the proposal of \cite{Gillis_JPA1} that will be described in detail in Section III they play the central role because it is assumed that it is interactions that actually induce the collapse. In other proposals they are crucial to establishing the large scale entanglement that allows nonlinear collapse to occur on macroscopic scales while leaving the (almost) linear quantum behavior of microscopic systems essentially undisturbed. Because these equations typically require an extremely large number of elementary systems to become entangled in order to generate collapse it is essential that the entanglement relations are well defined throughout the collapse process. To insure this there must be some means of sequencing the interactions that generate these relations. In typical measurement processes some of these interactions are spacelike-separated. The assumption of a preferred reference frame (or more generally, a foliation of spacetime) provides the necessary sequencing.

In addition to a foliation (or some similar structure) the other critical feature that must be introduced is a stochastic process (or processes). As mentioned earlier any nonlinear and nonlocal modification of the Schr\"{o}dinger equation must be stochastic in order to prevent superluminal signaling. The stochastic process to be described here is based on the Wiener integral of a white noise Gaussian process. This can be thought of as the continuous time limit of an unbiased random walk with zero mean. As such, it scales with time as $\sqrt{t}$. It is designated as $\xi(t)$, and its differential, which plays a key role in the equations, is designated as $d\xi(t)$. In general, $\xi(t)$ can be complex. The process is governed by the rules of the It$\hat{o}$ stochastic calculus\cite{Ito_51}: $ d\xi^{*} d\xi = dt, \;\;\; dt d\xi = 0.$
 
More general stochastic processes can be considered, and a number of proposals employ a stochastic field which is a function of both space and time, $\xi(x,t)$, rather than just time. There is also another form of the stochastic calculus due to Stratonovich. See the references for details. A good general reference is the text by Gardiner.\cite{Gardiner}

\subsection{Nonlinear, Norm-Preserving Stochastic Collapse Equations}

The overall aim of this work is to show how relativity and the nonlocal aspects of quantum theory can be encompassed in a unified mathematical structure describing the dynamics of physical systems without the need to introduce ad hoc rules that limit the applicability of the mathematics. However, the general form of the stochastic collapse equations outlined here (which includes the proposal described in Section III) is formulated in a nonrelativistic framework. The main reason for this is to avoid mathematical complexity. The assumption of a preferred reference frame that will be used here should be seen as just the simplest special case of a randomly evolving spacelike surface. The restriction to nonrelativistic quantum mechanics makes it possible to illustrate the essential ideas of the proposal in \cite{Gillis_JPA1} while avoiding many of the complications involved in quantum field theory. It also makes it possible to build on the substantial body of work developed in \cite{Gisin_c,Pearle_1976,Pearle_1979,Gisin_1984,GRW,Diosi_1,Diosi_2,Diosi_3,GPR,GP_1,GP_2,Adler_Brun, Ghirardi_Bassi,Pearle_1,Brody_finite}.

Measurements of quantum systems can have a very large number of possible outcomes, but at the most basic level they come down to either a detection or a failure to detect. The essentially binary character of measurement processes means that at each stage the Hilbert space (of any number of dimensions) can be decomposed into two orthogonal subspaces, and the state vector of the total system can be represented as the sum of two components, one in each subspace. Although the particular decomposition can vary during the process (thus, allowing any number of possible outcomes) the fundamentally binary nature makes it possible to model measurement processes as a random walk between two alternatives.\footnote{When   there are more than two possible outcomes the alternatives consist of two disjoint sets of outcomes, one of which will increase in amplitude and the other decrease.}  This is essentially what stochastic collapse equations do.\footnote{Wave function collapse can occur either inside or outside a laboratory, but for ease of explanation the discussion here will focus primarily on typical measurement situations.}

To see how these equations work we can begin with a simple theorem about the relative probabilities of a random walk ending at either of the two end points. If the walk continues for enough steps it will eventually finish at one or the other of the end points. Label the end points as $0$ and $1$, and suppose that the walk begins at a point, $p$, between them. It will be shown that the probability of ending at $1$ is $p$ and that the probability of ending at $0$ is $1-p$. The step size is labeled $\delta$. It is allowed to vary anywhere between $0$ and the distance to the nearest end point: 
      $ 0 \, \leq \, \delta \, \leq \, p $,  $ 0 \, \leq \, \delta \, \leq \, 1-p $. 
Label the probability of reaching $1$ as $Pr(p)$. Because the walk is assumed to be unbiased we get: 
   $ Pr(p) \, = \, \frac{1}{2} Pr(p-\delta) + \frac{1}{2} Pr(p+\delta)$. Because this relationship holds for all values of $p$ and $\delta$ it is linear. Therefore, the probability of reaching $1$ is $p$, and the probability of reaching $0$ is $1-p$. What follows is an attempt to present an intuitive explanation of how nonlinear, norm-preserving stochastic collapse equations map onto this simple, binary picture.

The binary character of measurement processes noted above allows us to illustrate the essential operation of stochastic collapse equations using a simple two state system. Consider a system with a wave function, $\psi $, and Hamiltonian, $\mathbf {H}$. The system evolves under the action of the Hamiltonian and of a stochastic collapse operator that is constructed from a self-adjoint operator, $\mathbf{O}$, with two eigenstates, $|x\rangle$ and $|y\rangle$, associated with eigenvalues $\mathbf{a}$ and $ \mathbf{b}$. The eigenstates of the operator, $\mathbf{O}$, define the collapse basis. The nonlinear stochastic operator is defined as:
$ \, \mathcal{O} \; \equiv \; k[ \mathbf{O} - \langle \, \psi | \mathbf{O}| \psi \, \rangle ] \,  $, where $\langle \, \psi | \mathbf{O}| \psi \, \rangle $ is the expectation value of the operator, $\mathbf {{O}}$, in the state, $\psi$, and $k$ is a constant that helps to determine the strength and scale of the collapse effects and also insures that $\mathcal{O}$ is dimensionless. For example, if $\mathcal{O}$ is based on the position operator $k$ could determine the range of the collapse effect.  

The modified stochastic Schr\"{o}dinger equation is defined as:
\begin{equation}\label{1N}    
d\psi \,   = \,  \frac{-i}{\hbar} \mathbf {{H}} \, \psi \, dt \,  + \, 
\mathcal{O}   \, \, \psi\,  \sqrt{\gamma} \, d\xi(t)  \, - \, \frac{1}{2}  \mathcal{O}^2  \, \psi \, \gamma \, dt .  
\end{equation}
The first term on the right represents the standard Schr\"{o}dinger evolution. The primary stochastic action is described by the middle term. The parameter, $\gamma$, determines the rate at which the stochastic operator acts. The square root operator is applied to the rate parameter, $\gamma$, because it works in conjunction with the stochastic differential, $d\xi$, which scales as $\sqrt{dt}$ (as described above). In the definition of the stochastic operator, $\mathcal{O} $, the subtraction of the expectation value, $\langle \, \psi | \mathbf{O}| \psi \, \rangle $, acts, as in the Gram-Schmidt procedure, to insure that the stochastic modification of $\psi$ is orthogonal to the existing wave function. The small orthogonal addition to the wave function slightly alters the norm. This alteration is compensated for by the third term on the right which involves $\mathcal{O}^2$.

The action of the stochastic term is quite small in comparison to that of the Hamiltonian. So, in order for it to be effective in generating collapse it needs to act in a manner that is essentially independent of the Hamiltonian. This can be achieved in several ways, through appropriate choices for the rate parameter, the operator, $\mathbf{O}$, and possibly other parameters.

 The stochastic term is designed to drive the system to one of the eigenstates of the operator, $\mathbf{O}$. In this simple example the wave function is represented as:
 \newline
 $\psi \, = \, \alpha |x\rangle + \beta |y\rangle$, with $ |\alpha|^2 \, + \, |\beta|^2 \, = \, 1$. To simplify the example these amplitudes can be taken as real and positive with no loss of generality. The action of the self-adjoint operator, $\mathbf{O}$, on the wave function is:  
 \newline 
$\mathbf{O} \psi \, = \,\mathbf{a} \alpha |x\rangle + \mathbf{b}\beta |y\rangle$, and its expectation value is 
$\langle \, \psi | \mathbf{O}| \psi \, \rangle \, = \,  \mathbf{a} \alpha^2 \, + \, \mathbf{b} \beta^2$. The action of the stochastic operator on the wave function can be expanded as:
\begin{equation}\label{2N}
\begin{array}{ll}    
\mathcal{O} \, \, \psi \,   = \, k\{ \mathbf{a} \alpha |x \rangle + \mathbf{b} \beta |y\rangle  - (   \mathbf{a} \alpha^2 \, + \, \mathbf{b} \beta^2 ) \big{[} \, \alpha |x\rangle +  \beta |y\rangle \, \big{]} \}
& \\  \, = \,k\{\ \alpha \, [ \mathbf{a} (1- \alpha^2) \, - \, \mathbf{ b} \beta^2 ] |x\rangle \,  
+ \, \beta \, [\mathbf{b} (1-\beta^2) \, - \, \mathbf{a} \alpha^2 ] |y\rangle \}
& \\  \, = \, k\{ \alpha \, \beta \, (\mathbf{a}-\mathbf{b}) \big{ [ } \beta |x\rangle \, -
\,  \alpha |y\rangle\big{ ]} \}.
\end{array}
\end{equation}
So the middle term of \ref{1N} can be written as: 
\begin{equation}\label{3N}
 k\{ \alpha \, \beta \;  \big{[}  \beta |x\rangle \, -
\,  \alpha |y\rangle \big{]}  \; (\mathbf{a}-\mathbf{b})  \sqrt{\gamma} \, d\xi(t) \}.  
\end{equation}

The expression in square brackets can be recognized as a normalized vector that is orthogonal to $\psi$. As long as the rate parameter, $\gamma$, is independent of $\alpha$ and $\beta$ the only dependence on the amplitudes (aside from the orthonormal state vector) is the term, $\alpha \beta$.

In this form and with the simplifying assumptions described above it is possible to trace the evolution of the wave function through Hilbert space under the influence of the stochastic operator. Since $\alpha$ and $\beta$ are assumed to be real and positive the evolution can be modeled as a random walk along the arc joining $x$ and $y$ axes (corresponding to the eigenstates $|x\rangle$ and $|y\rangle$).\footnote{In this simplified picture with $\alpha$ and $\beta$ real and positive the process can be pictured on a standard Cartesian graph with x and y axes. It is very straightforward to transfer the analysis to the Bloch sphere afterward.} The state, 
$\psi \, = \, \alpha |x\rangle \, + \, \beta |y\rangle$, lies on the arc; the orthogonal state,  $\beta |x\rangle \, - \, \alpha |y\rangle$, is tangent to the arc and it drives the state to one or the other of the eigenstates in infinitesimal steps. The magnitude and direction of the steps are determined by the coefficient of the tangent state vector, 
$   \alpha \, \beta \; k (\mathbf{a}-\mathbf{b})  \sqrt{\gamma} \, d\xi(t). $ \footnote{The role of the third term in equation \ref{1N} involving $ \mathcal{O}^2 $ is simply to keep the resulting state on the arc by maintaining the correct normalization.}

The dependence on $\alpha$ and $\beta$ means that the step size varies as the wave function, $\psi$, traverses the arc under the influence of the stochastic operator. 
Since the effective distance to the end points, $|x\rangle$ and $|y\rangle$, is measured in terms of this step size the distance calculation must take this variation into account. This can be done by introducing a parameter, $\theta$, with $\alpha \, = \, \cos \, \theta$ and $\beta \, = \, \sin \, \theta$, and integrating the coefficient of the tangent state vector along the arc:
\begin{equation}\label{4N}
k(\mathbf{a}-\mathbf{b})  \sqrt{\gamma} \, d\xi(t) \; \int  \, \cos \, \theta \, \sin \, \theta \, d\theta \; = \; A\sin^2 \, = A\beta^2.  
\end{equation}
where $A$ is some term independent of $\alpha, \beta$ and $\theta$. If we associate the two eigenstates with the end points of a random walk as described earlier, with $|x\rangle$ corresponding to $0$ and $|y\rangle$ corresponding to $1$, then the position of $\psi$ along the arc can be parameterized as $\beta^2$. As shown earlier, this is the probability that the random walk ends at $|y\rangle$, and $\alpha^2 \, = \, 1-\beta^2$ is the probability that it ends at $|x\rangle$. So the Born rule follows from the basic structure of the collapse equation in a straightforward manner. It is also worth noting that the random walk terminates because the term, $\alpha \beta \, = \cos \, \theta \, \sin \, \theta$ goes to $0$ as $\psi$ approaches one of the end points.\footnote{The fact that $\alpha \, \beta$ approaches $0$ at the end points also creates a problem (the ``tails problem") in that the walk does not end in a finite number of steps. This problem will not be dealt with in detail here, but I will offer a speculative solution later.}

I have assumed here that the terms, $k$ and $\gamma$, are constant, and, hence, independent of the amplitudes, $\alpha$ and $\beta$. This independence from the amplitudes is crucial for the derivation of the Born rule. It is this feature that must be maintained when $k$ and $\gamma$ are replaced by parameters that vary in a manner that is dependent on the interaction strength. These parameters are discussed in the next section.

As stated at the beginning of this section what has been shown here is that, if we are willing to countenance some additional fundamental structure for relativistic spacetime and incorporate the probabilistic character of quantum theory at the fundamental level it is possible to modify the Schr\"{o}dinger equation in a fairly simple way so that it yields the measurement postulates as dynamic consequences. In this way contemporary physical theory is rendered much more coherent.

With this background we can now review the proposed equation described in \cite{Gillis_JPA1}, and see how the stochastic modification maintains Lorentz invariance.

\section{Interaction-Induced Wave Function Collapse}

 Measurements consist of interactions that establish correlations between physical systems. Correlations are established through the exchange of conserved quantities. 
 Given the probabilistic nature of quantum theory, the generation of stable information  and its transmission depend on these correlating interactions. These considerations are what motivated and guided the construction of the equation described below. Roughly speaking, the idea is that the magnitude of the collapse effect associated with an interaction is proportional to the amount of correlation that is generated.

 The degree of correlation between two systems that is generated during an interaction depends on the extent to which the interaction changes the individual state of each system. This, in turn, depends on the strength of the interaction and the resistance of each of the systems to a change of state. The strength is measured by the interaction potential energy. This is assumed to depend on the separation between the two systems and to decrease as the separation, 
 $\mathbf{r} $, increases. For systems $j$ and $k$ it will be indicated as 
 $\mathbf {{ V}_{jk}} \, = \, \mathbf {{ V}_{jk}} (\mathbf {{r}_{jk}})$. The resistance to change depends on the mass of the systems, $m_j$ and $m_k$. The effective mass of elementary systems such as electrons in bound states is altered by the binding interactions. So atoms, molecules, and other complex structures are treated as single systems with a total mass and a net charge (or electric multipole moment).

 So the collapse operator is based on the interaction potential energies, $\mathbf{{V}_{jk}}$, and is proportional to the ratio, $\mathbf{{V}_{jk}} / (m_j+m_k)$. As mentioned in Section II the stochastic operator must be dimensionless. To convert the denominator to an energy it is multiplied by the square of the speed of light, $c^2$. 
 This is the only nonarbitrary speed, it is crucial for maintaining Lorentz invariance, and it eliminates the need to introduce an arbitrary constant, $k$. 
 It is also necessary, as in Section II, to subtract the expectation value, 
 $\langle \, \psi | \mathbf{ { V}_{jk}}| \psi \, \rangle $. 
So the component of the collapse operator associated with the interaction between systems $j$ and $k$ is:
\begin{equation}\label{5N}    
{\mathcal{V}}_{jk} \; \equiv \;  \frac{\mathbf{ { V}_{jk}} - \langle \, \psi | \mathbf{ { V}_{jk}}| \psi \, \rangle}  {(m_j+m_k)c^2 }.  
\end{equation}

In most proposed collapse equations the rate or frequency parameter, $\gamma$, is  a constant. It is chosen in a rather ad hoc manner to minimize deviations from linearity at an elementary level while insuring collapse on a macroscopic scale. In contrast, since it is assumed here that collapse effects are induced by the physical processes that establish correlations between systems it is possible to define this parameter in terms of the rate at which the correlations are generated. Since correlations are established through the exchange of conserved quantities, and since these exchanges are associated with variations in the interaction potentials, $\mathbf{ { V}_{jk}}$, we can define a rate parameter, $\gamma_{jk}$, associated with each interaction in terms of the rate at which the interaction proceeds. As described in Section II the term, $\sqrt{\gamma_{jk}} $, is multiplied by the stochastic differential, $d\xi$.

The rate can be defined so that it integrates to a value of order, $1$, over the course of the interaction, and goes to zero when the interacting systems either settle into a stationary state or separate to a distance at which the interaction effectively ends. This insures that the stochastic collapse equation reduces to the ordinary Schr\"{o}dinger equation in these situations. This can be done by taking the ratio of the rate of change of potential energy to the maximum magnitude of potential energy that occurs during the interaction:
\newline
 (\textit{rate of change of P.E. / max P.E}).

Both the rate of change and the maximum value are affected by the action of the Hamiltonian and the stochastic operator. However, because the stochastic action is unbiased and because the changes that it induces in the numerator and denominator are correlated its effects tend to cancel out. Because collapse equations involve stochastic differentials, $ d \xi $, time derivatives are not well defined.  But, it is possible to use the Hamiltonian to express its contribution to the rate of change of potential energy in terms of spatial derivatives:
  \begin{equation}\label{gamm_num} 
  \begin{array}{ll}            
  \Big{|} \Big{|} \; (i \hbar )  \, \mathbf{\int} \, \mathbf{ { V}_{jk}} \, \Big{(} \,     
  \frac{ \psi^* \, \mathbf{\nabla_j}^2   \psi  \, 
  	- \, \psi \mathbf{\nabla_j}^2  \psi^* } {2m_j} 
  \,  +  \,  
  \frac{ \psi^* \, \mathbf{\nabla_k}^2   \psi \, 
  	- \, \psi \mathbf{\nabla_k}^2    \psi^* } {2m_k} \, \Big{)} \; \Big{|} \Big{|} .
  \end{array}  
  \end{equation}	 
The norm is taken to insure a positive rate. 

To establish that the expression for the denominator in $\gamma_{jk}$, $|| \langle \, \psi |  \mathbf{ { V}_{jk}} | \, \psi \rangle ||_{max}$, is well defined note that the collapse equation can be integrated both backward and forward in time. Because the Hamiltonian is independent of time it is clear that the associated Schr\"{o}dinger equation is integrable. The fact that the stochastic operator is unbiased implies that averaging over all of the unravellings of the collapse equation will reproduce the value obtained by the Schr\"{o}dinger equation integral. Observe also that the stochastic changes in the denominator are matched by those in the numerator; so their overall effect on $\gamma_{jk}$ is to leave it essentially unchanged. It is, of course, not necessary to actually carry out the integration since $\gamma_{jk}$ is designed to integrate to a value of order, $1$, over the course of the interaction. With this understanding the variable rate parameter is defined as follows: 
	\begin{equation}\label{gamm_def}
	\gamma_{jk} \, \equiv  \, \frac{ ||  (i \hbar )  \, \mathbf{\int} \, \mathbf{ { V}_{jk}}  \Big{(}      
		\frac{ \psi^*  \mathbf{\nabla_j}^2   \psi   
			-  \psi \mathbf{\nabla_j}^2  \psi^* } {2m_j} 
		  +    
		\frac{ \psi^*  \mathbf{\nabla_k}^2   \psi  
			-  \psi \mathbf{\nabla_k}^2    \psi^* } {2m_k} \, \Big{)}  || } {|| \langle \, \psi |  \mathbf{ { V}_{jk}} | \, \psi \rangle ||_{max} }.
	\end{equation}

	An important point to note about \ref{gamm_def} is that both numerator and denominator pick out\textit{ only the interacting components of the the wave function}. This means that $\gamma_{jk}$ depends only on the portion of the wave function that is involved in the interaction, and thus the rate parameter is independent of the amplitudes of the interacting and noninteracting components. This feature is necessary to insure compliance with the Born probability rule as illustrated in the discussion in Section II.

 The period during which $\gamma_{jk}$ is significantly different from zero depends on the initial conditions of the interaction. However, the time during which the vast majority of the momentum and energy are exchanged (and the amplitude is transferred) depends essentially on the maximum interaction energy:
 \begin{equation}\label{9N}    
 dt_{int} \,  \approx \, \frac{\hbar} {   \mathbf{ {V}_{jk-max}}  },   
 \end{equation}
 For two electrons with a maximum interaction energy equal to the potential at the Bohr radius the duration would be about $2.5*10^{-17}$ seconds. Rates and durations for other interactions can be scaled from this estimate, taking into account the mass and charge of the systems involved.

As promised a few paragraphs back this formulation for $\gamma_{jk}$ implies that when multiplied by $dt$ it integrates to a value of order $1$ over the course of the interaction. This is also true for the expression $ \sqrt{\gamma_{jk}}d\xi$. This fact allows one to treat each interaction as a discrete event, and makes it possible to estimate the scale on which collapse occurs and the duration of the collapse process. This kind of analysis shows how the scale and duration depend on the average strength of the interactions involved. The estimates of the average interaction rate and duration were given above.

Finally, the full collapse operator is obtained by multiplying the operators by the square root of the rate parameters and summing over the terms for each $(j,k)$ pair: 
\begin{equation}\label{10N}    
{\mathcal{V}} \; \equiv \;   \sum_{j < k}  {\mathcal{V}}_{jk} \sqrt{\gamma_{jk}} .
\end{equation}
The proposed collapse equation takes the form:
 \begin{equation}\label{collapse_eqn}   
 \begin{array}{ll}  
 d\psi \,   = \,  (-i / \hbar) \mathbf {{H}} \, \psi \, dt \,  +\, 
 \sum_{j < k}  {\mathcal{V}}_{jk}   \, \, \psi\, \sqrt{\gamma_{jk}} d\xi(t)  
 & \\   
 \; \; \;\;\;\; \;\;\;\; \;\;\;\; \; \;  \;\;\;\; \; \;  \;\;\;\; \; \; \;\; \; \;- \, \frac{1}{2}  (\sum_{j < k}  {\mathcal{V}}_{jk})^2 \,  \psi \, \gamma_{jk} dt .  
 \end{array} 
 \end{equation}

 A detailed proof that \ref{collapse_eqn} results in collapse with the correct probabilities is given in \cite{Gillis_JPA1}, along with estimates of the scale and duration of collapse processes. As indicated these depend on the average strength of the interactions involved. Such processes can involve anywhere from about $10^8$ to $10^{16}$ elementary interactions. Since many of the interactions can be occurring in parallel the durations are typically very small fractions of a second as can be seen from the estimates of the durations associated with individual interactions given above.

 The claim that conservation laws hold \textit{exactly} in individual instances of collapse obviously runs counter to the prevailing presumption that conservation laws hold only on average in quantum theory. However, this presumption is based on an artificial division of the world into classical and quantum systems, and also on an overly idealized concept of elementary physical systems being in strictly factorizable states. In order to properly assess the status of conservation laws in quantum theory it is necessary to treat all systems, both macroscopic and microscopic, as quantum systems, and also to recognize that all these systems have a history of interaction with other systems (which include preparation apparatuses).  As emphasized in \cite{Gemmer_Mahler,Durt_1} interaction generates entanglement. Therefore, the idealization of elementary systems as being in purely factorizable states is never fully realized in practice. This is pointed out by the authors of \cite{Gemmer_Mahler} where they say:
\begin{quote}
	``Thus it is, strictly speaking, unjustified to describe a particle in a box,
	which is part of an interacting quantum system, by a wave-function”.
\end{quote}
In other words, the common textbook example of a particle in a box ignores the small amount of entanglement that results from the interaction between the particle, the box and whatever apparatus was used to prepare the system. The interactions involved in both the preparation and measurement of quantum systems are conservative interactions. When these interactions induce branching of the wave function they lead to a different distribution of conserved quantities among the interacting systems in the various branches, but they do not alter the total amount (up to normalization of the branches). Based on these kinds of observations a number of articles in recent years have demonstrated that conservation laws hold exactly in individual instances, and not just when applied to ensembles of identical measurement situations. In addition to \cite{Gillis_JPA1} these include \cite{Gillis_IJQF2,Gillis_IIWFCRCL,Zeng_Sun_Shao,Gillis_CQM,Marletto_Vedral,APR_2,APR_2_seq}. 

The collapse process involves the shift of amplitude among distinct branches of the wave function. Because the total amount of a conserved quantity up to normalization is the same in each branch these shifts do not, by themselves, violate any conservation law. 

Violations could occur if they are induced by the process that generates the shifts. There are a couple of ways in which such  violations can be induced for the general type of collapse equations considered here. One of these is illustrated by the most widely known collapse proposal, the continuous spontaneous localization (CSL) model\cite{GPR,Ghirardi_Bassi}, which implies small violations of energy conservation as described in \cite{CSL_Exp}.

The CSL proposal introduces two new constants and a stochastic \textit{field}, $\xi(x,t)$, which is a function of both position and time. The new constants together with the stochastic variations in both space and time act as a source of external energy that can be fed into other physical systems. In contrast, the proposed equation in \cite{Gillis_JPA1} does not introduce any new constants and it uses a single global stochastic process,  $\xi(t)$, which is a function only of time. Therefore, this problem does not arise.

Another potential problem concerns the use of a stochastic process based on white noise. Since white noise includes arbitrarily high frequencies it will introduce energy violations if it is not modulated. To avoid this a number of proposals have employed colored noise, in which the high frequencies are strongly attenuated. In the current proposal the stochastic differential, $d\xi(t)$, is multiplied by the variable interaction rate, $\sqrt{\gamma_{jk}}$. The term, $\sqrt{\gamma_{jk}} d\xi(t)$, is significantly different from zero for only a very brief time, and it heavily suppresses frequencies that are much different from  $\frac{\mathbf{ {V}_{jk-max}}}{\hbar}$ (the inverse of equation \ref{9N}). This prevents spurious high frequency noise from injecting additional energy into the system.

Explicit demonstrations of how the proposed equation maintains conservation laws in individual measurement situations are given in in \cite{Gillis_JPA1}. Momentum and angular momentum are conserved exactly. Because the proposal is formulated in a nonrelativistic framework it is only able to conserve energy within the accuracy allowed by the limited forms of energy describable in nonrelativistic theory. Specifically, it is shown that the proposed modification exactly conserves energy in those situations in which the Schr\"{o}dinger equation \textit{correctly} predicts energy conservation. In cases where the Schr\"{o}dinger equation fails to track relativistic adjustments to the nonrelativistic energy formulas the differences from those formulas  that the stochastic modification predicts match the relativistic corrections both qualitatively and quantitatively.

 Experimental consequences of the proposal are also discussed in the earlier work. These deal with very small discrepancies in the correlations between entangled systems predicted by conventional quantum theory and the equation described above. They are a result of the nonlinearity of the equation.

 This review of the proposal in \cite{Gillis_JPA1} is intended as background for the demonstration that the action of the stochastic operator in \ref{collapse_eqn} is Lorentz invariant. The argument is presented in the next section.

 \section{Lorentz Invariance}

 To establish Lorentz invariance it must be shown that the amplitude shift induced by the interaction of two physical systems is independent of the reference frame from which that interaction is viewed. As described in the previous section the magnitude of the amplitude shift is determined by the ratio of the interaction energy to the total relativistic energy of the interacting systems. The stochastic operator associated with the shift is active only during the very brief period during which conserved quantities are being exchanged. The rate parameter is determined by the rate of change of interaction energy, and when multiplied by the stochastic differential in the appropriate way it integrates to a value of order one over the course of the interaction. It will be shown here that both the ratio of the interaction energy to the total relativistic energy of the two systems and the integration of the interaction rate over the course of the correlating phase of the interaction are independent of the reference frame.

 The total relativistic energy of interacting systems, $j$ and $k$, consists of the rest energy and kinetic energy of the two systems, along with the interaction (aka ``potential") energy: 
\begin{equation}\label{tot_ener}    
 \Big{(} \frac{m_j}{\sqrt{1-\beta_j ^2}} + \frac{m_k}{\sqrt{1-\beta_k ^2}}  \Big{)} c^2 \, +\mathbf{ V_{jk}}.
\end{equation}
In this expression $m_j$ and $m_k$ represent the rest mass of the two systems.\footnote{In the earlier discussion the difference between the rest energy and total relativistic energy was ignored due to its very small value in nonrelativistic interactions.} The terms, $\beta_j$ and $\beta_k$, depend on the velocities of the two systems within the frame from which they are viewed. Relativistically, energy is transformed as the time component of the energy-momentum four-vector: 
\begin{equation}\label{ener_trans}    
	E' = \Gamma(E - u p_\parallel)
\end{equation}
where $E'$ is the energy in the primed frame, $u$ is the velocity of the primed frame relative to the unprimed frame, $p_\parallel$ is the parallel component of momentum of the system in the unprimed frame, and capital $\Gamma$ is the relativistic transformation expression, $ \frac{1}{\sqrt{1-\beta^2}}$ (where $\beta = u/c$). What must be shown is that the ratio of interaction energy, $\mathbf{ V_{jk}}$, to the total relativistic energy (given in \ref{tot_ener}) is the same in all frames. 

The transformation of interaction energy (represented as $\mathbf{ V_{jk}}$) in a particular case depends on whether it contributes to the momentum in the unprimed frame. For example, in equation \ref{ener_trans} above if $\mathbf{ V_{jk}}$ is not a factor in the expression for $p_\parallel$ then the transformation is given simply by multiplication by $\Gamma$. This follows immediately from \ref{tot_ener}) and \ref{ener_trans}. 

What we want to show is that, for an interaction between systems, $j$ and $k$, the ratio of the interaction energy to the total relativistic energy of the systems is the same, no matter what inertial frame the interaction is viewed from. In the rest frame of the center of mass of the interacting systems the total energy is given by equation \ref{tot_ener} above. Since $ \mathbf{p}_j  = - \mathbf{p}_k $ in the c-o-m frame the total momentum is zero. Therefore, the transformation of both the interaction energy and the  total relativistic energy from the c-o-m frame to \textit{any} other frame is given simply by multiplication by $\Gamma$. Thus, the ratio is the same in every frame.

  As described earlier the rate parameter associated with the stochastic operator,  $\gamma_{jk}$, is designed to integrate to a value of order, $1$, over the period during which correlations are being established through the exchange of conserved quantities: 
  \begin{equation}\label{rate_x_time}
  \; \int  \gamma_{jk}  dt \; \sim \; 1.  
  \end{equation}
  The duration of the exchange as viewed from different reference frames varies according to the usual relativistic formula. It is clear that the rate at which such an interaction proceeds must vary inversely with the duration.  Hence, the integral of the rate over the course of an interaction must be the same for all reference frames. According to the It$\hat{o}$ calculus rules the variance of the stochastic process over a time, $\Delta t$, is $\Delta t$:
 \begin{equation}\label{pref_stoch_int}
 \; \int_t^{t+\Delta\,t}  \, d\xi(s) d\xi^{*}(s)  ds \; = \; \Delta \, t.  
 \end{equation}  
 Over the period of the interaction \ref{rate_x_time} and \ref{pref_stoch_int} imply that 
\begin{equation}\label{int_root}
\; \int   \sqrt{\gamma_{jk}} d\xi   \; \sim \; 1.  
\end{equation}

It is also clear that transformations of the rate parameter, $\gamma_{jk} $, from one frame to another must vary in a synchronized manner relative to ${\mathcal{V}}_{jk}$. Therefore, the two key elements of the collapse term, ${\mathcal{V}}_{jk}$  and $\gamma_{jk}$, have the correct Lorentz transformation properties.\footnote{It is true that the rate parameter, $\gamma_{jk}$, has been defined using the divergence of the current density based on the nonrelativistic Hamiltonian. However, for the low interaction energies and relative kinetic energies this yields a value essentially equivalent to the relativistic formula. Moreover, it varies inversely with respect to time according to the usual relativistic transformation formula.} So, if we look at the primary stochastic term,
\begin{equation}\label{collapse_term}   
{\mathcal{V}}_{jk}   \, \, \psi\, \sqrt{\gamma_{jk}} d\xi,    
\end{equation}
everything transforms properly with the exception of the nonrelativistic version of the wave function, $\psi$. To complete the argument concerning Lorentz invariance we need to look at the entanglement relations that underlie the nonrelativistic approximation.

The fundamental idea behind this proposal is that wave functions are induced to collapse to one of their branches by the entangling interactions that constitute measurements. It is entanglement relations that define the branches and mediate collapse; \textit{the entanglement established by previous interactions is crucial to the collapse process}.

This proposal was developed in a nonrelativistic framework in which the $j$ and $k$ systems are pictured as interacting instantaneously at spacelike separations. In reality the causal influences that generate the interactions propagate at finite speeds over timelike or lightlike intervals. In the simplified account based on the nonrelativistic framework both ends of an instantaneous $j-k$ interaction are multiplied by the same ${\mathcal{V}}_{jk}\sqrt{\gamma_{jk}}d\xi$ term, and this is critical both in maintaining conservation laws and insuring the overall consistency of the collapse process. This is no longer true in the more realistic case involving finite propagation speeds; so we need to see how entanglement operates to maintain consistency.

Consider an interaction between systems $j$ and $k$. It consists of a sequence of causal influences emitted by one of the systems and the reaction of the other system to those influences. Suppose at time, $t_1$, system $j$ reacts to an influence emitted at a slightly earlier time by system $k$ and emits a causal influence. This interaction is accompanied by a stochastic change in the wave function proportional to ${\mathcal{V}}_{jk}\sqrt{\gamma_{jk}}d\xi \,(t_1)$. The causal influence carries with it the effects of the stochastic operator at that time. At a slightly later time, $t_2$, system $k$ reacts to the influence from system $j$ and emits a causal influence. This interaction is also accompanied by a stochastic change proportional to ${\mathcal{V}}_{jk}\sqrt{\gamma_{jk}}d\xi \,(t_2)$. Because of the entanglement that is generated by the interaction the effects of the stochastic operator acting at time, $t_2$, propagate back to system $j$. So both ends of the interaction are subject to the same \textit{two} stochastic changes. The interaction proceeds in this ``ping-pong" fashion. This ensures the consistency of the collapse process. The simplified, instantaneous interaction description gives a condensed, approximate version of this process.

This analysis can now be used to generate a quantitative estimate of the accuracy of the nonrelativistic approximation involving $\psi$.  Because this proposal was formulated in a nonrelativistic framework the interaction energies are assumed to be small relative to the total relativistic energy of the systems involved. These interaction energies are typical of those used in a very wide range of measurements on quantum systems. Although the  $j-k$ system can have a very high energy in the preferred frame, the low interaction energies imply that in the center-of-mass frame the relevant velocities are nonrelativistic. Over the fairly small spacetime region in which the interaction occurs one can transform to the c-o-m frame and analyze the interaction with nonrelativistic quantum mechanics.

There are two primary factors that can affect the accuracy of the nonrelativistic approximation. The first of these  stems from the fact that the nonrelativistic wave function, $\psi$, is defined on a different family of spacelike hypersurfaces from that of the preferred frame. The second concerns the extent to which the $j-k$ components of the wave function, $\psi$, can evolve during the propagation delays. To evaluate the magnitude of these effects note first that the large majority of the amplitude transfer occurs close to the point at which the interaction energy reaches its maximum. This can be easily verified 
from the definitions of ${\mathcal{V}}_{jk}$  and $\gamma_{jk}$. Typical nonrelativistic measurement interactions take place over small distances and short times; fairly typical interactions can occur at $10^{-10}$ meters and $10^{-16}$ seconds. So we need to assess how large a fractional deviation from the correct amplitudes could be induced over this range.

 To determine the maximum possible temporal discrepancy between various portions of the interacting $j-k$ wave function associated with the hyperplane variation we can divide the interaction distance by the speed of light:  $ \frac{10^{-10} } { 3*10^8 } \; \sim \; 10^{-18}$ seconds. Since the causal influences discussed above generally propagate at the speed of light discrepancies resulting from propagation delays are of the same magnitude. 
 The maximum speeds for systems interacting at this (minimum) distance are about $10^6$ meters per second. So the maximum possible discrepancies in the positions of the various interacting portions of the systems are about $10^{-12}$ meters. This is about $0.01$ of the separation of the systems. This fraction provides a measure of accuracy for the approximation. 
 
 The considerations above apply to any interaction independent of the relationship of the preferred frame to the center-of-mass frame for that particular interaction. We have seen that the term, $\sqrt{\gamma_{jk}}d\xi$, integrates to a value of order $1$ no matter which frame is the preferred one. It is clear, therefore, that, independent of which reference frame is preferred, the average magnitude of the amplitude shift will be of order, 
 \begin{equation}\label{5N_b}    
 {\mathcal{V}}_{jk} \; \equiv \;  \frac{\mathbf{ { V}_{jk}} - \langle \, \psi | \mathbf{ { V}_{jk}}| \psi \, \rangle}  {(m_j+m_k)c^2 }.  
 \end{equation} 
The direction and specific magnitude of the amplitude transfers vary from case to case, but this is determined by the random nature of the stochastic process - not by the selection of a preferred frame. Thus the amplitude shifts governed by \ref{collapse_eqn} are Lorentz invariant.

In Section II it was shown that stochastic collapse equations provide a very natural extension of the mathematical formalism of quantum theory and that they entail the measurement postulates as straightforward consequences of the fundamental equation. This general approach eliminates the need to insert them into the logical structure of the theory at some very vaguely defined point. In Section III it was shown that the particular modification of the Schr\"{o}dinger equation proposed in \cite{Gillis_JPA1} insures the maintenance of conservation laws in individual measurement processes without introducing any new, ad hoc physical constants.   In this section it was shown that the collapse operator is Lorentz invariant. The next section considers what implications this demonstration has for our understanding of the relationship between nonlocal quantum effects and relativity.

\section{Relativity, Nonlocal Quantum Effects, and Spacetime Ontology}

The principle of relativity states that the laws of physics have the same form in all reference frames. Since the speed of light has been observed to be independent of the reference frame, a satisfactory theory should explain this invariance in a natural way without positing any unnecessary features.

Early works by Fitzgerald, Lorentz, Poincar{\'e}, and Larmor that were aimed at explaining the observed invariance assumed that electromagnetic waves propagated through a stationary aether. These researchers accounted for the experimental results by proposing that their measurement instruments - rods, clocks, and all matter - interacted according to the laws of electromagnetism which had been shown to be Lorentz invariant. The resulting behavior of the instruments made it impossible to pick out any special reference frame. Thus, Lorentz invariance became the principal defining characteristic of a relativistic theory. In his review of these early works Bell\cite{Bell_Rel} illustrates how they established that 
\begin{quote}
	``if physical laws are Lorentz invariant...moving observers will be unable to detect their motion."
\end{quote}
The term, ``motion", of course, refers to motion with respect to the aether. What Einstein was able to show was that the hypothesis of an undetectable aether was superfluous; it added nothing to the explanation of the observed phenomena.

In the previous section it was shown that the stochastic modification to the  Schr\"{o}dinger equation proposed in \cite{Gillis_JPA1} is Lorentz invariant, and that, therefore, the preferred frame that it assumes remains undetectable. What then distinguishes this situation from the pre-Einstein versions of relativity? The critical difference is that in this case the hidden spacetime feature plays a critical explanatory role. It provides a framework within which the basic dynamic equation of the theory can \textit{explain} the nonlocal correlations described by Bell. Moreover, the fundamentally probabilistic nature of the collapse equation also explains why the preferred frame remains hidden.

What does this imply about our understanding of relativity? Following Einstein's 1905 papers \cite{Einstein_1,Einstein_2} Minkowski presented his elegant mathematical description\cite{Minkowski} which provided a very compelling account of spacetime structure. The Lorentzian metric defined a framework in which it appeared that all physical processes were constrained to propagate only within light cones. This provided an explanation for Einstein's postulates in terms of the fundamental features of spacetime. The fact that it dovetailed beautifully with our intuitive notion that causal processes propagate
through space in a continuous manner was very satisfying. When Einstein was able to generalize this picture to account for gravitational effects without invoking action-at-a-distance it cemented the idea that we had a complete account of spacetime ontology.

The advent of quantum theory radically altered this situation. Einstein was the first to recognize this, but it was Bell's demonstration of the reality of nonlocal effects that has really forced us to reconsider the status of relativity and the implications for the ontology of spacetime. In classical theory Lorentz invariance followed from Einstein's two postulates, but this is no longer true in quantum theory. As emphasized by Weinberg, the assumption of local commutativity is \textit{essential} to maintain this defining characteristic of a relativistic theory. It is time to acknowledge that it constitutes a third postulate of relativity, and that, like Einstein's two original postulates it should be explained in terms of spacetime structure. The nonlocal correlations should, of course, be explained in a natural way. The stochastic collapse equation described in Section III shows how this can be done.

 The adoption of the additional stochastic structure described in Section III transfers the probabilistic nature of quantum theory from the macroscopic to the elementary level. This turns what were ad hoc postulates at the macro level into straightforward consequences of the mathematical description of elementary processes. The probabilistic nature of elementary correlating interactions also explains why a preferred reference frame (or spacetime foliation) remains hidden, and explains why the description of spacetime at the macroscopic level is limited to the standard relativistic account involving just a Lorentzian metric. 
 
 It is important to note that although the additional spacetime structure posited by stochastic collapse equations remains hidden there is, nevertheless, experimental evidence for its existence. The numerous experiments confirming the nonlocal correlations identified by Bell strongly argue for some kind of connection across spacelike intervals. Additional confirmation is possible. The proposals offered to explain the correlations usually make some testable predictions. For example, the proposal in \cite{Gillis_JPA1} predicts specific quantitative deviations from the linearity of standard quantum theory. These deviations are, of course, quite small. They are proportional to the square of the amplitude shifts, that is the ratio of interaction energy to total relativistic energy of the interacting systems. For two electrons separated by the Bohr radius these deviations from linearity would be about $10^{-9}$. Relevant experiments would be quite challenging, but they are not impossible.

 Because the proposal described here is formulated in a nonrelativistic framework it must be considered as incomplete. But since the proposed modification is Lorentz invariant there do not appear to be any serious conceptual obstacles to a full relativistic account. Such an account would encompass quantum field theory, and might well offer solutions to some of the key problems with this more limited proposal. As argued in \cite{Gillis_JPA1} there is reason to believe that such an extension could resolve the small discrepancies with energy conservation that are implied by equation \ref{collapse_eqn}. One might also hope that a relativistic account might provide a solution to the tails problem (mentioned in a footnote in section II). With the transition from distinguishable to indistinguishable particles the suppression of the very small amplitudes in the wave function tail below the level of vacuum fluctuations might point the way to a resolution. These possibilities will be addressed in future work. 
 \newline
 \newline
 \textbf{Acknowledgements:} 
	I would like to thank Nicolas Gisin and Flavio Del Santo for helpful discussions.

\end{document}